\documentstyle[12pt]{article}

\def\boxit#1{
\vbox{\hrule height0.5pt\hbox{\vrule width0.5pt\kern10pt\vbox{
\kern10pt#1\kern10pt}
\kern10pt\vrule width0.5pt}\hrule height0.5pt}}

\def\bild#1\over#2{\mathrel{\mathop{\kern5pt #1}\limits_{#2}}}

\newcommand{\be}{\begin{equation}}
\newcommand{\ee}{\end{equation}}
\newcommand{\bc}{\begin{center}}
\newcommand{\ec}{\end{center}}
\newcommand{\ba}{\begin{array}}
\newcommand{\ea}{\end{array}}

\begin{document}

\title{ Response Function of Hot Nuclear Matter }
\author{F. L. Braghin\thanks{Doctoral fellow of Coordena\c c\~ao de
 Aperfei\c coamento de Pessoal de N\'\i vel Superior,Brasil }
  and D. Vautherin \\
Division de Physique Th\'eorique\thanks{Unit\'e de Recherche des
Universit\'es Paris XI et Paris  VI associ\'ee au CNRS},\\
 Institut de Physique Nucl\'eaire, 91406, Orsay Cedex, France
}
\date{}
\maketitle

{\bf Abstract}:   We investigate the response function of hot nuclear matter to
a small isovector external field using a simplified Skyrme
interaction reproducing the value of the symmetry energy coefficient.
 We consider values of the momentum transfer corresponding to the
dipole oscillation in heavy nuclei. We find that while at zero temperature the
 particle hole  interaction is almost repulsive enough to have a sharp
(zero sound type)
collective oscillation, such is no longer the case at temperatures of a few
MeV. As a result a broadening of the dipole resonance occurs, leading to its
quasi disappearence by the time the temperature reaches
5 MeV. The sensivity of the temperature evolution of the width when modifying
the residual interaction strength is also examined.
\vspace*{4cm}

\noindent IPNO/TH 94-15\hfill{June  1994}

\newpage
\setcounter{equation}{0}
  A large amount of data has been accumulated over the years concerning the
properties of giant resonances in hot nuclei \cite{GAA92,NVM93} and the
evolution of their
energy and width as temperature increases. To a good accuracy the energy of
giant dipole resonance is found to be independent of the excitation energy
while its width is found first to increase at low excitations energies
with a saturation
at higher energies \cite{BRA89}. On the theoretical side the energy of the
giant dipole resonance is well understood already at the level of the random
phase approximation (RPA). Concerning  the increase of the width with T
however,
the situation is still unsettled. Indeed although there are various mechanisms
contributing to this increase such as shape fluctuations or angular momentum
effects \cite{GAL85,ALH90} their total contribution does not seem to account
for
 the observed
data. Furthermore there are also dynamical effects inhibiting this increase
\cite{BRO87}.
 It is worthwhile noting that in some of the early RPA calculations using
oscillator functions \cite{VV84}  a rapid increase of the width was found. This
result was however not confirmed in the self-consistent RPA calculations
 including continuum states performed by Sagawa and Bertsch \cite{SAG84}.
Recently the early RPA calculations have been reexamined by Nicole Vinh Mau
\cite{NVM92} who concluded that one important ingredient in these
calculations was the
appearence of new particle-hole configurations as temperature rises, an effect
to which  more attention should be given when trying to explain the broadening
of the giant dipole resonance. Why are these effects found to be nearly
negligible in the calculations of Sagawa and Bertsch? The aim of the present
letter is to examine this question in the framework of a self consistent
calculation of the response function of nuclear matter using a schematic Skyrme
force. The advantage of this calculation is that while incorporating a full
self-consistency it is still tractable and leads to transparent formulae.
In particular these formulae show that while at zero temperature the isovector
 particle -hole residual  interaction is almost repulsive enough to have a well
developped zero sound, such is no longer the case at temperatures of a few
MeV, leading in general to a rapid weakenig of collective effects.

In the case of nuclear matter single particle states are labeled by their wave
number {\bf k}, their spin $\sigma$ and their isospin $\tau$.
 We wish to calculate the response to an external field of the form:
\be \label{1}
V_{ext}= \epsilon \tau_3 e^{-i {\bf q}.{\bf r} } e^{-i (\omega +i \eta) t}.
\ee
where $\tau_3$ is the third component of the isospin operator and $\eta$ a
vanishingly  small positive number corresponding to an adiabatic switching of
the external field. For this purpose we use mean field theory. In this case
the evolution of the one-body density matrix $\rho$ is determined by the time
dependent Hartree-Fock equation:
\be \label{2}
i \partial_t \rho= [W + V_{ext},  \rho ],
\ee
where W is the mean field hamiltonian. Let us consider for simplicity the case
of a schematic Skyrme force:
\be \label{3}
v= t_0(1+ x_0 P_\sigma) \delta ({\bf r}_1 -{\bf r}_2) + t_3 \delta
({\bf r}_1 -{\bf r}_2) \delta ({\bf r}_2 -{\bf r}_3) .
\ee
In this case we have
\be \label{3a}
W = \frac{p^2}{2m} + U(r),
\ee
where $U$ is a local
potential, given in the case of neutrons by:
\be \label{4}
U_n({\bf r},t) = \frac{t_0}{2} \{ 2 \rho({\bf r},t)- \rho_n({\bf r},t) \} +
\frac{t_3}{4} \{\rho^2({\bf r},t)- \rho_n^2({\bf r},t) \},
\ee
with a similar expression for protons. In the previous equation
 $\rho({\bf r},t)$ is
the density distribution $<{\bf r}| \rho |{\bf r} >$.   An identical equation
would be obtained with a density dependent version of the Skyrme force
 \cite{RS82} and thesubsequent developpement would be unmodified.
 For a small enough external field
it is legitimate to linearize the mean field evolution equation (\ref{2}). This
procedure leads to the following approximate equation for the difference
$\delta \rho = \rho_n - \rho_p$ of the neutron and proton density matrices:
\be \label{5}
\ba{ll}
i \partial_t <{\bf k} | \delta \rho | {\bf k}'>=
&\{ \epsilon({\bf k})- \epsilon({\bf k}') \} <{\bf k}| \delta \rho| {\bf k}'>
+ \{ f({\bf k}')-f({\bf k}) \} <{\bf k}| (U_n-U_p)| {\bf k}'> \\
&+2 \epsilon \{ f({\bf k}')-f({\bf k}) \}
\delta({\bf k}'-{\bf k}- {\bf q}) e^{-i (\omega +i \eta) t} .
\ea
\ee
In this equation $ \epsilon ({\bf k})= \hbar^2 k^2/2m$
 is the energy of the
single particle state with wave number ${\bf k}$ and
\be \label{5a}
 f({\bf k})=1/\{1+e^{\beta(\epsilon({\bf k})-\mu )} \},
\ee
is the corresponding
occupation number. Note that in the linear response approximation we have
from equation (\ref{4}):
\be \label{6}
U_n({\bf r}, t)- U_p({\bf r},t)= 2 V_0 \delta \rho({\bf r},t)
\ee
where
\be \label{7}
V_0=- \frac{t_0}{2} (x_0+ \frac{1}{2}) - \frac{t_3}{8} \rho_0,
\ee
$\rho_0$ being the saturation density of nuclear matter.
Note that $V_0$ is related to the symmetry energy coefficient $a_{\tau}$
of nuclear matter via the relation:
\be \label{7a}
a_{\tau}= \frac{1}{3} T_F+ \frac{1}{2} V_0 \rho_0.
\ee
where $ T_F $ is the kinetic energy at the Fermi surface.
The above equations
suggest to look for a solution of the evolution equation (\ref{5}) in the form:
\be \label{8}
<{\bf k}| \delta \rho (t) | {\bf k}'>= \delta ({\bf k}'- {\bf k}-{\bf q})
<{\bf k}| \delta \rho (t=0) | {\bf k}+ {\bf q}> e^{-i (\omega +i \eta) t}.
\ee
For this Ansatz the change in the density distribution has the following
structure:
\be \label{9}
 \delta \rho ({\bf r},t)=
\alpha e^{-i {\bf q}.{\bf r} } e^{-i (\omega +i \eta) t},
\ee
where:
\be \label{10}
\alpha= \sum_{\bf k} <{\bf k}| \delta \rho (t=0) | {\bf k}+ {\bf q}> .
\ee
 The change between the neutron and proton potentials is thus found from
equation (\ref{6}) to be:
\be \label{11}
<{\bf k}| (U_n-U_p)| {\bf k}'> = 2 V_0 \alpha \delta ({\bf k}'- {\bf k}+{\bf
q})
e^{-i (\omega +i \eta) t}.
\ee
Inserting this expression into the linearized evolution equation for the
density matrix (\ref{5}) we find that our Ansatz is indeed a solution
provided that:
\be \label{12}
<{\bf k}| \delta \rho(0)| {\bf k}+ {\bf q}> = \frac
{ (2 V_0 \alpha+2 \epsilon) \{ f({\bf k})- f( {\bf k}+ {\bf q}) \} }
{\omega +i \eta-  \epsilon ({\bf k}) + \epsilon({\bf k}+ {\bf q})  }.
\ee
Returning to equation (\ref{10}) we obtain a linear relation determining the
value of
$\alpha$:
\be \label{13}
\alpha= (V_0 \alpha + \epsilon) \Pi_{0R}(\omega,{\bf q}),
\ee
In this equation $\Pi_{0R}$ is the unperturbed retarded response function
defined by:
\be \label{14}
\Pi_{0R}(\omega, {\bf q})= \frac{2}{ (2 \pi)^3}
\int d {\bf k} \frac
{ f({\bf k} + {\bf q})- f( {\bf k})  }
{\omega +i \eta-  \epsilon ({\bf k}) + \epsilon({\bf k}+ {\bf q})  },
\ee
Solving equation (\ref{13}) for $\alpha$ we obtain the following
expression for the
retarded response function $\Pi_R$ in the RPA approximation:
\be \label{15}
\Pi_R(\omega, {\bf q}) \equiv  \frac{\alpha}{\epsilon}= \frac{\Pi_{0R}
(\omega,{\bf q}) }{1- V_0 \Pi_{0R}(\omega, {\bf q})  }.
\ee
The result we have derived is valid only in the case of a simplified Skyrme
force. For other interactions the quantity $V_0$ would not be a c-number. The
case of a full Skyrme force was worked out at zero temperature by
Garcia-Recio et al \cite{GAR92} who still obtained a closed expression for the
response function at the cost of introducing additional unperturbed functions.
For a given momentum {\bf q} we find that there is a resonant response
when the frequency $\omega$ corresponds to a zero of the denominator, i.e.,
 when:
\be \label{16}
1= V_0  \Pi_{0R} (\omega, {\bf q}) .
\ee
The real part of $\omega$ determines the energy of the collective mode while
$\Im m(\omega)$ determines its life time \cite{WAL71}. The importance of the
resonance condition (\ref{16}) is also seen when looking at the distribution
of strength per unit volume $S(\omega)$ for the operator
$\exp(i {\bf q}. {\bf r})$. Indeed it is given by $-\Im m \Pi_R / \pi$ i.e.
\cite{WAL71}
\be \label{19}
S(\omega) = -  \frac{1}{\pi} \frac{\Im m \Pi_{0R}(\omega, q)}
{ (1 - V_0 \Re e\Pi_{0R})^2 + (V_0 \Im m \Pi_{0R})^2}.
\ee
For symmetric nuclear matter this function enjoys the following sum rule
\be \label{19a}
\int_0^{\infty} \omega S(\omega) d \omega= \frac{\hbar^2}{2m} \rho_0 q^2.
\ee
 Let us now try to apply the previous formulae to the case of finite nuclei.
For this purpose a usefull guide is the Steinwedel and Jenssen model
\cite{RS82}.
In this model neutrons and protons oscillate inside a sphere of radius $R$
according to the formula:
\be \label{17}
\rho_n({\bf r},t) -\rho_p({\bf r},t)= \varepsilon
\sin ({\bf q}.{\bf r}) \exp(i \omega t),
\ee
the total density remaining equal to the saturation density $\rho_0$ and
the wavenumber $q$  being given by
\be \label{18}
q = \frac{\pi}{2 R}.
\ee
In order to describe the dipole resonance in finite nuclei this model thus
suggests to look at the strength function $S(\omega,q)$ calculated at a
momentum transfer $q=\pi/2R$, where $R$ is the nuclear radius.
In the following we shall focus on the nucleus lead-208 for which
we take
$R = 6.7 fm $ and $ q=.23 fm^{-1} $. We  use the following values of the
 parameters:
$ t_{0} = -983.4$ ${\rm MeV} \times {\rm fm}^3$, $t_{3} = 13106 $ ${\rm MeV}
\times {\rm fm}^6 $ and $ x_{0} = .48 $.
 These values
were fitted to reproduce the binding energy  (E/A= -16MeV),
the saturation density( $\rho_0$ = 0.17 fm$^{-3}$) and the symmetry
energy($a_{\tau} =$ 30 MeV) of nuclear matter. For these values $V_0=203 MeV
fm^3$.

 To calculate S($\omega$) we still need the expression of the unperturbed
Lindhard function. The imaginary part can be expressed in terms of elementary
 functions as \cite{BCV90}:
\be \label{20}
 \Im m \Pi_{0R}(\omega , q) = -\frac{k T}{8 \pi q} \left(\frac{2m}{\hbar^{2}
}\right)^{2} \log \frac{1+ e^{\beta (A+ \omega /2)}}{1+e^{\beta(A- \omega/2)}}
\ee
where $ \beta = 1/kT$ while
\be \label{21}
 A=\mu-\frac{\omega^{2}}{4 \epsilon(q)} - \frac{\epsilon(q)}{4}.
\ee
 In this equation $\mu$ is the chemical potential and $\epsilon(q)$ the single
particle energy of the state with momentum $q$. The real part can be expressed
as:
\be \label{22}
\Re e \Pi_{0R} = - \int F(k) df(k)
\ee
where $f(k)$ is the occupation number while:
\be \label{23}
F(k)= \frac{1}{2 \pi^{2}} \frac{mk}{\hbar^2} \left\{ -1 +
\frac{k}{2q}\left[ \phi( \frac{mw}{\hbar kq}+ \frac{q}{2k})-
\phi( \frac{mw}{\hbar kq}-\frac{q}{2k})\right] \right\}
\ee
with:
\be \label{24}
\phi (x)= (1-x)(1+x) \log | \frac{x-1}{x+1} |.
\ee
 Note that at zero temperature we have $ df(k) = -\delta (k-k_F)dk $ so that
$F(k)$ is just  the zero temperature value of the real part of $ \Pi_{0R} $.
This
implies that equation (\ref{22}) is well suited for a numerical integration
centered
around the value $k = \sqrt{ 2 m \mu / \hbar^2}$.

The real part of the Lindhard function is graphed on figure 1 as a function of
the frequency $\omega$ for $q=.23 fm^{-1}$ and several values of the
temperature T=0,2,4,6 MeV. It can be seen that at T=0 the quantity
$\Re e \Pi_{0R}$ has
 a maximum near $\omega = 13 MeV $ which is not quite sufficient to produce a
zero in the quantity $ 1 - V_0 \Re e \Pi_{0R} $. Such a zero would
correspond to a sharp collective oscillation analogous to the zero sound mode
in neutral Fermi liquid systems \cite{WAL71}. Still, for T=0, the value of
 $ V_0 \Re e\Pi_{0R} $
is sufficiently close to unity at its maximum to produce a peak in
the strength at $\omega = 14 MeV $, as seen in figure 2. It should be noted
that the position of this peak is in close agreement with the observed value
(14 MeV). It is also interesting to note in figure 1 that as temperature
increases the maximum in $ \Re e \Pi_{0R} $ becomes weaker and weaker as a
result of the averaging with the occupation numbers in equation 26. This leads
to a broadening of the peak in the response function as can be checked in
figure 2. By the time one reaches a temperature of 6 MeV only small collective
effects remain visible.

The analysis we have just given also explains why some RPA calculations do give
a rapid increase of the dipole width and some others do not. Indeed a slightly
stronger residual interaction would be able to produce at zero temperature a
sharp
zero sound type mode which would disappear rapidly as temperature increases. In
contrast a weaker residual interaction gives at zero temperature an already
broad peak evolving slowly with temperature. An illustration of this point
is given
in the lower part of figure 2, which shows the evolution of the dipole mode for
a residual interaction with $ V_0$ = 100 MeV $\times$ fm$^3 $.
Similar results are obtained for lighter nuclei i.e. larger values of the
momentum transfer $q$. One difference however is that the maximum of the real
part of the response function becomes less pronounced as $q$ increases which
produces broader resonances.  Of course the discussion we have given ignores
finite size or surface effects and is further limited to the RPA framework.
Some effects  not included in this framework,  such as the coupling to two
particle-two hole states, are known to be important to describe
the damping of collective vibrations \cite{ADA84}.
Our analysis was further based on the Steinwedel and Jenssen model.  We believe
that analogous results would be obtained with the Goldhaber Teller model
\cite{RS82}.
 Indeed we expect in this case the response function to be dominated by the
$ q = \pi/ 2 R$ mode with small corrections only arising from the multiples
$ q = n \pi/ 2 R$ ($n > 1$),
 because the peak in $ \Re e \Pi_{0R}$ is much smaller for $n>1$.

{\bf Acknowledgements} We are grateful to Nguyen Van Giai and Nicole Vinh Mau
for stimulating discussions.

\newpage
\bc
{\bf Figure Captions}
\ec
{\bf Figure 1} Real part of the non perturbed Lindhard function multiplied
 by $\hbar c$ (in fm$^{-2}$) as a  function
of the energy $\omega$ (in MeV) for a momentum q=.23 $fm^{-1}$ and for
different values of the temperature T=0,2,4 and 6 MeV.

\noindent {\bf Figure 2} Distribution of strength per unit volume for the
operator
$\exp (i {\bf q}. {\bf r})$ (in MeV$^{-1} \times$ fm$^{-3}$) as a function
of the energy $\omega$ (in MeV) for a momentum q=.23 fm$^{-1}$ and for
different values of the temperature T=0,2,4 and 6 MeV. The upper curves
correspond
to an interaction strength $V_0$= 203 MeV$\times$fm$^3$ while the lower curve
corresponds to $V_0$= 100 MeV $\times$ fm$^3$.

\end{document}